\input{epsf}
\documentclass[11pt]{article}
\def\be{\begin{equation}}
\def\eea{\end{eqnarray}}
\def\bea{\begin{eqnarray}}
\def\ee{\end{equation}}

\author{M.Mohammadi$^{1,2}$ \footnote{majid471702@yahoo.com} , M.H.Naderi$^{3}$ \footnote{mhnaderi2001@yahoo.com} and M.Soltanolkotabi$^{3}$ \footnote{soltan@sci.ui.ac.ir}
\\ $^{1}$ {\small Department of Physics, Science and Research Campus Azad University of Tehran, Tehran, Iran}
\\$^{2}$ {\small Department of Physics, Shahreza Islamic Azad University, Shahreza, Isfahan, Iran}
\\$^{3}$ {\small Quantum Optics Group, Department of Physics, University of Isfahan, Isfahan, Iran}}
\title{Quantum statistical properties of the Jaynes-Cummings
           model in the presence of a homogeneous gravitational field }
 
\begin{document}
\maketitle
\begin{abstract}
\noindent The temporal evolution of quantum statistical properties
of an interacting atom-field system in the presence of a
homogeneous gravitational field is investigated within the
framework of the Jaynes-Cummings model. Taking into account both
the atomic motion and gravitational field a full quantum
treatment of the internal and external dynamics of the atom is
presented based on an alternative su(2) dynamical algebraic
structure. By solving analytically the Schr\"{o}dinger equation
in the interaction picture, the evolving state of the system is
found by which the influence of the gravitational field on
 the dynamical behavior of the atom-field system is explored. Assuming that initially the
field is prepared in a coherent state and the two-level atom is
in a coherent superposition of the excited and ground states, the
influence of gravity on the atomic dipole moment, collapses and
revivals of the atomic motion, atomic momentum diffusion, photon
counting statistics and quadrature squeezing of the radiation
field is studied.
\end{abstract}
\noindent PACS numbers: $42.50.VK,42.50.DV$\\
{\bf Keyword}: Jaynes-Cummings model, atomic motion,
gravitational field, Non-classical properties\\
\\
\section{Introduction}
The interaction between a two-level atom and  a single quantized
mode of the electromagnetic field in a lossless cavity within the
rotating wave approximation (RWA)can be described by the
Jaynes-Cummings model (JCM) [1]. Despite being simple enough to
be analytically soluble in the RWA, this model has been a
long-lasting source of insight into the nuances of the
interaction between light and matter. The JCM has been applied to
investigate many quantum effects such as the quantum collapses
and revivals of atomic inversion [2], squeezing of the radiation
field [3] ,atomic dipole squeezing [4], vacuum Rabi oscillation
[5] and the dynamical entangling and disentangling of the
atom-field system in the course of time [6-9]. Investigations of
the dynamical behavior of the JCM are also extremely important
due to its experimental realizations in high-Q microwave cavities
[10], in optical resonators [11], in laser-cooled trapped ions
[12] and in quantum nondemolition measurements [13]. Stimulated
by the success of the JCM, more and more people have paid special
attention to extending and generalizing the model in order to
explore new quantum effects. Discussions related to several
interesting generalizations of this model are now available in
the literature [14] and the model is still promising in many
applications, particularly in the tast developing research area
of quantum information [15].\\
\hspace*{00.5 cm} A very significant and noteworthy
generalization of the JCM is to include the effect of atomic
motion so that the spatial mode structure could be incorporated
into this model. In the standard JCM, the interaction between a
constant electric field and a stationary (motionless) two-level
atom is considered. With the development in the technologies of
laser cooling and atom trapping the interaction between a moving
atom and the field has attracted much attention [16-25]. In
particular, it has been shown that the atomic motion can bring
about the nonlinear transient effects similar to self-induced
transparency (SIT) and adiabatic following (AF) [26], the
possibility of realizing an optical switching [25], change the
creating time of Schr\"{o}dinger cat states [22] and exhibit long
time entropy squeezing effect [24].\\
\hspace*{00.5 cm} On the other hand, experimentally, atomic beams
with very low velocities are generated in laser cooling and
atomic interferometry [27]. It is obvious that for atoms moving
with a velocity of a few millimeters or centimeters per second
for a time period of several milliseconds or more, the influence
of Earth's acceleration becomes important and cannot be neglected
[28]. For this reason it is of interest to study the temporal
evolution of a moving atom simultaneously exposed to the
gravitational field and a singel-mode traveling wave field. Since
any quantum optical experiment in the laboratory is actually made
in a non-inertial frame it is important to estimate the influence
of Earth's acceleration on the outcome of the experiment.
Recently, a semiclassical description of a two-level atom
interacting with a running laser wave in a gravitational field
has been studied [29,30]. In Ref.[31] a complementary scheme based
on an su(2) dynamical algebraic structure to investigate the
influence of the gravity on the QND measurement of
atomic momentum in the dispersive JCM has been studied.\\
\hspace*{00.5 cm} In this paper we adopt a dynamical algebraic
approach to investigate the temporal evolution of quantum
statistical properties of the JCM in the presence of a
homogeneous gravitational field. In the Jaynes-Cummings model,
when the atomic motion is in a propagating light wave, we
consider a two-level atom interacting with the quantized
cavity-field in the presence of a homogeneous gravitational
field. By solving analytically the Schr\"{o}dinger equation in
the interaction picture, the evolving state of the system is
found by which the influence of the gravitational field on
 the dynamical behavior of the atom-field system is explored.
 In section 2, we present a full quantum treatment of the
internal and external dynamics of the atom with an alternative
su(2) dynamical algebraic structure within the system. Based on
this su(2) structure and in the interaction picture, we obtain an
effective Hamiltonian describing the atom-field interaction in
the presence of a gravitational field. In section 3 we
investigate the dynamical evolution of the system and show that
how the gravitational field may affect the dynamical properties
of the JCM. In section 4 we study the influence of gravitational
field on both the cavity-field and the atomic properties.
Considering the field to be initially in a coherent state and the
two-level atom in a coherent superposition of the ground and
excited states, we investigate the temporal evolution of the
atomic dipole moment, atomic inversion, atomic momentum
diffusion, probability distribution of the cavity-field, photon
counting statistics and quadrature squeezing of the radiation
field. Finally, we summarize our conclusions in section 5.
\section{ Jaynes-Cummings Model in the presence of Gravitational Field}
The system we consider here is a moving two-level atom of mass M
exposed simultaneously to a single-mode travelling wave field and
a homogeneous gravitational field. Therefore, the Hamiltonian of
the atom-field system  in the presence of gravitational field with
the atomic motion along the position vector $\hat{\vec{x}}$ and
in the rotating wave approximation is given by
\begin{eqnarray}
\hat{H}=&&\frac{\hat{p}^{2}}{2M}-M\vec{g}.\hat{\vec{x}}+\hbar\omega_{c}(\hat{a}^{\dag}\hat{a}+\frac{1}{2})+\frac{1}{2}\hbar\omega_{eg}\hat{\sigma}_{z}+\nonumber\\
&&\hbar\lambda[\exp(-i\vec{q}.\hat{\vec{x}})\hat{a}^{\dag}\hat{\sigma}_{-}+\exp(i\vec{q}.\hat{\vec{x}})\hat{\sigma}_{+}\hat{a}],
\end{eqnarray}
where $\hat{a}$ and $\hat{a}^{\dag}$ denote, respectively, the
annihilation and creation operators of a single-mode traveling
wave with frequency $\omega_{c}$, $\vec{q}$ is the wave vector of
the running wave and $\hat{\sigma}_{\pm}$ denote the raising and
lowering operators of the two-level atom with electronic levels
$|e\rangle, |g\rangle $ and Bohr transition frequency
$\omega_{eg}$. The atom-field coupling is given by the parameter
$\lambda$ and
 $\hat{\vec{p}}$, $\hat{\vec{x}}$
denote, respectively, the momentum and position operators of the
atomic center of mass motion and $g$ is Earth's gravitational
acceleration. It has been shown [31] that based on su(2) algebraic
structure, as the dynamical symmetry group of the model, the
Hamiltonian (1) can be transformed to the following effective
Hamiltonian
\begin{equation}
\hat{\tilde{H}}=\frac{\hat{p}^{2}}{2M}-\hbar
\hat{\triangle}(\hat{\vec{p}},\vec{g})\hat{S}_{0}+\frac{1}{2}Mg^{2}t^{2}+\hat{\vec{p}}.\vec{g}
t+\hbar (\kappa
\sqrt{\hat{K}}\hat{S}_{-}+\kappa^{*}\sqrt{\hat{K}}\hat{S}_{+}),
\end{equation}
where $\hat{\kappa}(t)$ is an effective coupling coefficient
\begin{equation}
\hat{\kappa}(t)= \lambda
\exp(\frac{it}{2}(\hat{\triangle}(\hat{\vec{p}},\vec{g})+\frac{\hbar
q^{2} }{M})),
\end{equation}
and the operators
\begin{equation}
\hat{S_{0}}=\frac{1}{2}(|e \rangle \langle e|-|g \rangle \langle
g|) , \hat{S_{+}}=\hat{a}|e\rangle \langle
g|\frac{1}{\sqrt{\hat{K}}},
\hat{S_{-}}=\frac{1}{\sqrt{\hat{K}}}|g\rangle \langle
e|\hat{a}^{\dag},
\end{equation}
with the following commutation relations
\begin{equation}
[\hat{S_{0}},\hat{S_{\pm}}]=\pm
\hat{S_{\pm}},[\hat{S_{-}},\hat{S_{+}}]=-2\hat{S_{0}},
\end{equation}
are the generators of the su(2) algebra and the operator
\begin{equation}
\hat{\triangle}(\hat{\vec{p}},\vec{g})=\omega_{c}-(\omega_{eg}+\frac{\vec{q}.\hat{\vec{p}}}{M}+\vec{q}.\vec{g}t+\frac{\hbar
q^{2}}{2M}),
\end{equation}
has been introduced as the Doppler shift detuning at time $t$
[31]. The Hamiltonian (2) has the form of the Hamiltonian of the
JCM, the only modification being the dependence of the detuning
on the conjugate momentum and the gravitational field. In the
interaction picture the transformed Hamiltonian (2) takes the
following form
\begin{equation}
\hat{\tilde{H}}_{int}=\exp(\frac{-i\hat{\tilde{H}}_{0}t}{\hbar})\hat{\tilde{H}}_{I}\exp(\frac{i\hat{\tilde{H}}_{0}t}{\hbar}),
\end{equation}
where
\begin{equation}
\hat{\tilde{H}}_{0}=-\hbar
\hat{\triangle}(\hat{\vec{p}},\vec{g})\hat{S}_{0}
+\hat{H}(\hat{\vec{p}},\vec{g}),
\end{equation}
and
\begin{equation}
\hat{\tilde{H}}_{I}=\hbar(\kappa
\sqrt{\hat{K}}\hat{S}_{-}+\kappa^{*}\sqrt{\hat{K}}\hat{S}_{+}),
\end{equation}
with
\begin{equation}
\hat{H}(\hat{\vec{p}},\vec{g})=\frac{\hat{p}^{2}}{2M}+\hat{\vec{p}}.\vec{g}t+\frac{1}{2}Mg^{2}t^{2}.
\end{equation}
Therefore we obtain
\begin{equation}
\hat{\tilde{H}}_{int}=\hbar (\hat{\kappa}(t)
\sqrt{\hat{K}}\hat{S}_{-}
\exp(-it\hat{\triangle}(\hat{\vec{p}},\vec{g}))+\hat{\kappa}^{*}(t)\sqrt{\hat{K}}\hat{S}_{+}\exp(it\hat{\triangle}(\hat{\vec{p}},\vec{g}))).
\end{equation}
 Finally by using Eq.(3) we arrive at
\begin{equation}
\hat{\tilde{H}}_{int}=\hbar \lambda(\sqrt{\hat{K}}\hat{S}_{-}
\exp(-it\hat{\triangle}_{1}(\hat{\vec{p}},\vec{g},t))+\sqrt{\hat{K}}\hat{S}_{+}\exp(it\hat{\triangle}_{1}(\hat{\vec{p}},\vec{g},t))).
\end{equation}
where
\begin{equation}
\hat{\triangle}_{1}(\hat{\vec{p}},\vec{g},t)=\frac{1}{2}(\omega_{c}-(\omega_{eg}+\frac{\vec{q}.\hat{\vec{p}}}{M}+
 \vec{q}.\vec{g}t+ 3\frac{\hbar q^{2}}{2M})).
\end{equation}
is the detuning of the atom-field interaction which depends on
both the atomic momentum and the gravitational field.
\section{Dynamical Evolution}
In section 2, we obtained an effective Hamiltonian for the
atom-field system in the presence of a homogeneous gravitational
field in the interaction picture. In this section, we investigate
dynamical evolution of the system. We will show how the
gravitational field may affect the quantum dynamics of JCM. For
this purpose, we solve the Schr\"{o}dinger equation
\begin{equation}
i \hbar \frac{\partial |\psi \rangle } {\partial
 t}=\hat{\tilde{H}}_{int}|\psi\rangle ,
\end{equation}
for the state vector $|\psi(t)\rangle$ with the Hamiltonian (12).
Indeed, the two-level atom with momentum $|\vec{p}\rangle$ in the
excited state $|e\rangle$ get annihilated and creates a field
excitation. Therefore, the Hamiltonian $\hat{\tilde{H}}_{int}$
transforms the state $|e\rangle \otimes |n\rangle \otimes
|\vec{p}\rangle \equiv |e,n\rangle \otimes |\vec{p}\rangle$, where
$|n\rangle$ denotes the nth Fock state of the field, into
\begin{equation}
\hat{\tilde{H}}_{int} |e,n\rangle \otimes |\vec{p}\rangle =
\hbar\lambda
\sqrt{n+1}\exp(-it\hat{\triangle}_{1}(\vec{p},\vec{g},t))|g,n+1\rangle
\otimes |\vec{p}\rangle,
\end{equation}
in which we have used the relations
\begin{equation}
\sqrt{\hat{K}}\hat{S}_{-}|e,n\rangle=\sqrt{n+1}|g,n+1\rangle ,
\hat{\vec{p}}|\vec{p}\rangle=\vec{p}|\vec{p}\rangle.
\end{equation}
Similarly, atom with momentum $|\vec{p}\rangle$ in the ground
state $|g\rangle$ get excited at the expense of annihilation a
field excitation. Hence, the Hamiltonian transforms the state
$|g\rangle \otimes |n+1\rangle \otimes
|\vec{p}\rangle\equiv|g,n+1\rangle \otimes|\vec{p}\rangle$ into
\begin{equation}
\hat{\tilde{H}}_{int} |g,n+1\rangle \otimes|\vec{p}\rangle =
\hbar\lambda
\sqrt{n+1}\exp(it\hat{\triangle}_{1}(\vec{p},\vec{g},t))|e,n\rangle
\otimes |\vec{p}\rangle.
\end{equation}
Since the Hamiltonian couples only the states $|g,n+1\rangle
\otimes |\vec{p}\rangle$ and $|e,n\rangle \otimes|\vec{p}\rangle$
we introduce the state vector
\begin{eqnarray}
|\psi(t)\rangle=&&\int d^{3}p
\sum_{n=0}(\psi_{e,n}(\vec{p},\vec{g},t)|e,n\rangle \otimes
|\vec{p}\rangle +
\psi_{g,n+1}(\vec{p},\vec{g},t)|g,n+1\rangle \otimes|\vec{p}\rangle) \nonumber \\
+&&\int d^{3}p \psi_{g,0}(\vec{p},t)|g,0\rangle
\otimes|\vec{p}\rangle.
\end{eqnarray}
The state  $|g,0\rangle$ which corresponds to $n=-1$ in Eq.(17)
plays a special role. According to Eq.(17) we find
$\hat{\tilde{H}}_{int}|g,0\rangle=0$ which means, the vacuum
cannot excite an atom initially in the ground state and
therefore, the state $|g,0\rangle$ decouples from the rest of the
states. Now we find the equations of motion for the time-dependent
probability amplitudes $\psi_{e,n}(\vec{p},\vec{g},t)=\psi_{1}$,
$\psi_{g,n+1}(\vec{p},\vec{g},t)=\psi_{2}$ by substituting (18)
into (14) and making use of Eqs.(15) and (17)
\begin{equation}
\dot{\psi}_{1}=-i\lambda
\sqrt{n+1}\exp(i\triangle_{1}(\vec{p},\vec{g},t)t)\psi_{2},
\end{equation}
and
\begin{equation}
\dot{\psi}_{2}=-i\lambda
\sqrt{n+1}\exp(-i\triangle_{1}(\vec{p},\vec{g},t)t)\psi_{1}.
\end{equation}
At time $t=0$ the atom is uncorrelated with the field  and the
state vector of the system can be written as a direct product
\begin{eqnarray}
|\psi(t=0)\rangle = &&|\psi_{c.m}(0)\rangle
\otimes|\psi_{atom}(0)\rangle \otimes|\psi_{field}(0)\rangle
\nonumber
\\=&& (\int d^{3}p
\phi(\vec{p})|\vec{p}\rangle)\otimes(c_{e}|e\rangle
+c_{g}|g\rangle )\otimes(\sum_{n=0}w_{n}|n\rangle),
\end{eqnarray}
where we have assumed that initially the field is in a coherent
superposition of Fock states, the atom is in a coherent
superposition of its excited and ground states, and the wave
vector for the center-of-mass degree of freedom is
$|\psi_{c.m}(0)\rangle=\int d^{3}p \phi(\vec{p})|\vec{p}\rangle$.
In notation (17) the initial state (21) reads
\begin{eqnarray}
|\psi(t=0)\rangle=&& \int d^{3}p
\sum_{n=0}(w_{n}c_{e}\phi(\vec{p})|e,n\rangle
\otimes|\vec{p}\rangle +
w_{n+1}c_{g}\phi(\vec{p})|g,n+1\rangle \otimes|\vec{p}\rangle)\nonumber \\
+ && \int d^{3}p w_{0}\phi(\vec{p})c_{g}|g,0\rangle \otimes
|\vec{p}\rangle.
\end{eqnarray}
When we compare (22) with (18) we find the initial conditions
\begin{equation}
\psi_{1}(t=0)=w_{n}c_{e}\phi(\vec{p})
,\psi_{2}(t=0)=w_{n+1}c_{g}\phi(\vec{p}),
\psi_{g,0}(t=0)=w_{0}c_{g}\phi(\vec{p}).
\end{equation}
We can solve two coupled first order differential equations (19)
and (20) in a straightforward way. We have
\begin{equation}
\frac{\partial^{2}\psi_{1}}{\partial
t^{2}}+2i\vec{q}.\vec{g}(t-\frac{\triangle_{0}}{2\vec{q}.\vec{g}})\frac{\partial\psi_{1}}{\partial
t}+\lambda^{2} (n+1)\psi_{1}=0,
\end{equation}
and
\begin{equation}
\frac{\partial^{2}\psi_{2}}{\partial
t^{2}}-2i\vec{q}.\vec{g}(t-\frac{\triangle_{0}}{2\vec{q}.\vec{g}})\frac{\partial\psi_{2}}{\partial
t}+\lambda^{2} (n+1)\psi_{2}=0,
\end{equation}
where
\begin{equation}
\triangle_{0}(\vec{p})=\frac{1}{2}[\omega_{c}-(\omega_{eg}+\frac{\vec{q}.\vec{p}}{M}+3\frac{\hbar
q^{2}}{2M})].
\end{equation}
is time-independent. Now, we solve analytically these equations
and we obtain
\begin{equation}
\psi_{1}(t)=\exp(i\triangle_{1}t)(C(1)H(A_{n},B_{t})+C(2)
_{1}F_{1}(-A_{n},\frac{1}{2};B_{t}^{2})),
\end{equation}
and
\begin{equation}
\psi_{2}(t)=C(1)H(A_{n}+1,B_{t})+C(2)
_{1}F_{1}(-\frac{1}{2}(A_{n}+1),\frac{1}{2};B_{t}^{2}),
\end{equation}
where $C(1)=\frac{C_{1}}{C}$,$C(2)=\frac{C_{2}}{C}$ with
\begin{equation}
C_{1}=\psi_{1}(0)_{1}F_{1}(-\frac{1}{2}(A_{n}+1),\frac{1}{2};(-D)^{2})-\psi_{2}(0)_{1}F_{1}(-A_{n},\frac{1}{2};(-D)^{2}),
\end{equation}
and
\begin{equation}
C_{2}=\psi_{1}(0)H(A_{n}+1,-D)-\psi_{2}(0)H(A_{n},-D),
\end{equation}
so that
\begin{equation}
C=H(A_{n},-D)_{1}F_{1}(-\frac{1}{2}(A_{n}+1),\frac{1}{2};(-D)^{2})-H(A_{n}+1,-D)_{1}F_{1}(-A_{n},\frac{1}{2};(-D)^{2}),
\end{equation}
and we have $A_{n}=-(2+i\beta)$,
$\beta=\frac{\Omega_{n}(\vec{p},\vec{g})-\Delta_{0}^{2}}{2\vec{q}.\vec{g}}$,
$B_{t}=(\gamma t-\eta)(1+i )$,
$\gamma=\frac{\sqrt{2}}{2}\vec{q}.\vec{g}$,
$\eta=\frac{\sqrt{2}\Delta_{0}}{4\sqrt{\vec{q}.\vec{g}}}$,
$D=\eta(1+i)$.
 We define $\Omega_{n}(\vec{p},\vec{g})=\sqrt{\Omega_{n}(\vec{p},0)^{2}+2i\vec{q}.\vec{g}}$
with
$\Omega_{n}(\vec{p},0)^{2}=\lambda^{2}(n+1)+\triangle_{0}^{2}$ as
the gravity-dependent Rabi frequency and $H(A_{n},B_{t})$,
$_{1}F_{1}(-A_{n},\frac{1}{2};B_{t}^{2})$ as the Hermite and the
hypergeometric functions, respectively.
\section{Dynamical Properties of The Model}
In this section, we study the influence of the gravitational
field on the quantum statistical properties of the atom and the
 quantized radiation field.  \\
\\
{\bf  4a. Atomic Dipole Moment} \\
\\
  When a two-level atom interacts with the cavity-field, a dipole moment is induced between the two atomic
levels. This induced dipole moment is given by the expectation
value of the dipole moment operator
\begin{equation}
P(t)=\langle\psi(t)|e\hat{\vec{x}}|\psi(t)\rangle.
\end{equation}
Therefore, from (18) we obtain
\begin{equation}
P(t)= \int
d^{3}p\sum_{n=0}^{\infty}[\psi^{*}_{1}\psi_{2}\wp_{eg}+\psi
^{*}_{2}\psi _{1}\wp^{*}_{eg}],
\end{equation}
where
\begin{equation}
\wp_{eg}=\wp^{*}_{ge}=e\langle
e|\hat{\vec{x}}|g\rangle=|\wp_{eg}|\exp(i\varphi),
\end{equation}
is the dipole matrix element and $\varphi$ is its phase. We
assume at $t=0$, atom is in a coherent superposition of the
excited state and the ground state $c_{g}(0)=\frac{1}{\sqrt{2}}$,
$c_{e}(0)=\frac{1}{\sqrt{2}}$, $\varphi=0$. We now consider
gravitational influence on the dipole moment evolution when in
$t=0$, the cavity-field is initially prepared in a coherent state
$w_{n}(0)=\frac{\exp(-\frac{|\alpha|^{2}}{2})\alpha^{n}}{\sqrt{n!}}$.\\
\\
In this case, by substituting (27) and (28) into (33) we have\\
\begin{eqnarray}
P(t)=&& 2|\wp_{eg}|\int d^{3}p |\phi(\vec{p})|^{2}
\sum_{n=0}^{\infty} Re \{ [ |C(1)|^{2}H(A_{n},B_{t})\\ \nonumber
&& H^{*}(A_{n}+1, B_{t}) + C(1)C^{*}(2)
H(A_{n},B_{t})_{1}F^{*}_{1}(\frac{-1}{2}(A_{n}+1),\frac{1}{2};B^{2}_{t})\\
\nonumber + && C(2)C^{*}(1)
H^{*}(A_{n}+1,B_{t})_{1}F_{1}(-A_{n},\frac{1}{2};B^{2}_{t})\\
\nonumber + &&
|C(2)|^{2}_{1}F_{1}(-A_{n}),\frac{1}{2};B^{2}_{t})_{1}F^{*}_{1}(\frac{-1}{2}(A_{n}+1),\frac{1}{2};B^{2}_{t})
] \exp(it\Delta_{1}(\vec{p},\vec{g},t))\},
\end{eqnarray}
where
$C_{\vec{p}_{0}}(1)=\frac{C_{1,\vec{p}_{0}}}{C_{\vec{p}_{0}}}$,
$C_{\vec{p}_{0}}(2)=\frac{C_{2,\vec{p}_{0}}}{C_{\vec{p}_{0}}}$
with
\begin{eqnarray}
 C_{1,\vec{p}_{0}}=&&\frac{\exp(-\frac{|\alpha|^{2}}{2})\alpha^{n}}{\sqrt{2(n!)}}(_{1}F_{1}(\frac{-1}{2}(A_{n}(\vec{p}_{0})+1),\frac{1}{2};D^{2}(\vec{p}_{0}))\\
\nonumber -&&(\frac{\alpha}{\sqrt{n+1}}) _{1}
F_{1}(-A_{n}(\vec{p}_{0}),\frac{1}{2};D^{2}(\vec{p}_{0}))),
\end{eqnarray}
and
\begin{eqnarray}
C_{2,\vec{p}_{0}}=&&\frac{\exp(-\frac{|\alpha|^{2}}{2})\alpha^{n}}{\sqrt{2(n!)}}(H(A_{n}(\vec{p}_{0})+1,-D(\vec{p}_{0}))\\
\nonumber -&&(\frac{\alpha}{\sqrt{n+1}})
H(A_{n}(\vec{p}_{0}),-D(\vec{p}_{0}))),
\end{eqnarray}
so that
\begin{eqnarray}
C_{\vec{P}_{0}}=&&H(A_{n}(\vec{p}_{0}),-D(\vec{p}_{0}))_{1}F_{1}(-\frac{1}{2}(A_{n}(\vec{p}_{0})+1),\frac{1}{2};D^{2}(\vec{p}_{0}))\\
\nonumber
-&&H(A_{n}(\vec{p}_{0})+1,-D(\vec{p}_{0}))_{1}F_{1}(-A_{n}(\vec{p}_{0}),\frac{1}{2};D^{2}(\vec{p}_{0})),
\end{eqnarray}
and we have $A_{n}(\vec{p}_{0})=-(2+i\beta_{\vec{p}_{0}} )$
$\beta_{\vec{p}_{0}}
=\frac{\Omega_{n}(\vec{p_{0}},\vec{g})-\Delta_{0}(\vec{p}_{0})^{2}}{2\vec{q}.\vec{g}}$,
$B_{t}(\vec{p}_{0})=(\gamma t-\eta_{\vec{p}_{0}})(1+i )$,
$\gamma=\frac{\sqrt{2}}{2}\vec{q}.\vec{g}$,
$\eta_{\vec{p}_{0}}=\frac{\sqrt{2}\Delta_{0}(\vec{p}_{0})}{4\sqrt{\vec{q}.\vec{g}}}$,
$D_{\vec{p}_{0}}=\eta_{\vec{p}_{0}}(1+i)$.
 Figure 1a show the dipole moment evolution assuming
$q=10^{7}m^{-1}$,
$p_{0}=10^{-26}\frac{Kg.m}{s}$,$g=9.8\frac{m}{s^{2}}$,$\omega_{rec}=.5\times10^{6}\frac{rad}{s}$,$\lambda=9.7\times10^{6}\frac{rad}{s}$,$\triangle_{0}=8.5\times10^{7}\frac{rad}{s}$
and $ \varphi=0$ [30-33]. Here we consider a two-level atom in a
coherent superposition of the excited state and the ground state
traversing in horizontal direction with the momentum vector
$\vec{p_{0}}$ of an optical cavity in the presence of gravitatonal
field so that $\vec{p_{0}}.\vec{g}=0$ and
$\vec{p_{0}}.\vec{q}=p_{0}q\cos\theta,
\vec{q}.\vec{g}=qg\sin\theta$ where $\theta$ is  the angle
between $\vec{q}$ and $\vec{p_{0}}$, and $\frac{\pi}{2}-\theta$ is
the angle between $\vec{q}$ and $\vec{g}$. Before a given atom
passes through the cavity, the cavity mode is prepared in the
coherent state. In figure 1b we consider small gravitational
influence.This means very small $\vec{q}.\vec{g}$, i.e., the
momentum transfer from the laser beam to the atom is only
slightly altered by the gravitational acceleration because the
latter is very small or nearly perpendicular to the laser beam.
With comparing figures 1a
and 1b we can see gravitational influence on the dipole moment by appearing oscillations such as collapses and revivals.\\
\\
\\
{\bf  4b. Atomic Inversion} \\
\\
Another important quantity is the  atomic population inversion
[34] which is given by the expression
\begin{equation}
w(\vec{p},\vec{g},t)=\langle\psi(t)|\sigma_{z}|\psi(t)\rangle,
\end{equation}
where from (18) we obtain
\begin{equation}
w(\vec{p},\vec{g},t)=\sum_{n=0}\int d^{3}p
[|\psi_{1}|^{2}-|\psi_{2}|^{2}].
\end{equation}
Therefore, by substituting from (27) and (28) into (40) and with
 assumes which we have used in the sub-section 4a, we can
obtain
\begin{eqnarray}
w(\vec{p}_{0},\vec{g},t)=&&\sum_{n=0}^{\infty}\{|C_{1,\vec{p}_{0}}|^{2}[
|H(A_{n}(\vec{p}_{0}),B_{t}(\vec{p}_{0}))|^{2} \\
\nonumber - && |H(A_{n}(\vec{p}_{0})+1,B_{t}(\vec{p}_{0}))|^{2} ]+
|C_{2,\vec{p}_{0}}|^{2}[|_{1}F_{1}(-A_{n}(\vec{p}_{0}),\frac{1}{2};B^{2}_{t}(\vec{p}_{0}))
|^{2}\\
\nonumber   - &&| _{1}F_{1}(- \frac{1}{2}(
A_{n}(\vec{p}_{0})+1),\frac{1}{2};B^{2}_{t}(\vec{p}_{0}))|^{2}] +
2Re[C_{1,\vec{p}_{0}}C^{*}_{2,\vec{p}_{0}}\\
\nonumber &&(H(A_{n}(\vec{p}_{0})+1,B_{t}(\vec{p}_{0}))
_{1}F^{*}_{1}(-A_{n}(\vec{p}_{0}),\frac{1}{2};B^{2}_{t}(\vec{p}_{0}))\\
\nonumber  -&& H(A_{n}(\vec{p}_{0})+1,B_{t}(\vec{p}_{0}))
_{1}F^{*}_{1}(- \frac{1}{2}(
A_{n}(\vec{p}_{0})+1),\frac{1}{2};B^{2}_{t}(\vec{p}_{0}))  ) ] \},
\end{eqnarray}
where we have defined all functions in terms of $\vec{p}_{0}$ in
the sub-section 4a.
 Figures 2a and
2b have plotted with the same corresponding data, respectively,
used in figures 1a and 1b.
 The gravitational field affect in the inversion population by appearing collapse and revival times so that we can see in figure
 2a. In figure 2b we consider $\vec{q}.\vec{g}=0$ so that we can
 not see collapse and revival times as well as figure 2a.\\
 \hspace*{00.5 cm}On the other hand, we calculate the
collapse and revival times [35-38]. We show that these times
depend on the gravitational field. An estimate of $t_{c}$ and
$t_{r}$ can be therefore be obtained from the conditions
\begin{equation}
(\Omega_{\langle n\rangle+\sqrt{\langle
n\rangle}}-\Omega_{\langle n\rangle-\sqrt{\langle
n\rangle}})t_{c}\sim 1,
\end{equation}
and
\begin{equation}
(\Omega_{\langle n\rangle}-\Omega_{\langle n\rangle-1})t_{r}\sim
2m \pi (m=1,2,3,...),
\end{equation}
with rabi frequency
\begin{equation}
\Omega_{n}=(c^{2}_{n}+d^{2})^{\frac{1}{4}}\exp(\frac{i\varphi_{n}}{2}),
\end{equation}
where
\begin{equation}
\tan(\varphi_{n})=\frac{d}{c_{n}},c_{n}=\triangle^{2}_{0}+\lambda^{2}(n+1),d=2\vec{q}.\vec{g}.
\end{equation}
Therefore,we obtain real part of the collapse and revival times
\begin{equation}
t_{c}=\frac{r_{1c}\cos(\varphi_{1c})-r_{2c}\cos(\varphi_{2c})}{(r_{1c}\cos(\varphi_{1c})-r_{2c}\cos(\varphi_{2c}))^{2}+(r_{1c}\sin(\varphi_{1c})-r_{2c}\sin(\varphi_{2c}))^{2}},
\end{equation}
and
\begin{equation}
t_{r}=\frac{2m\pi
r_{1r}\cos(\varphi_{1r})-r_{2r}\cos(\varphi_{2r})}{(r_{1r}\cos(\varphi_{1r})-r_{2r}\cos(\varphi_{2r}))^{2}+(r_{1r}\sin(\varphi_{1r})-r_{2r}\sin(\varphi_{2r}))^{2}},
\end{equation}
where
\begin{equation}
r_{1c}=(c^{2}_{\langle n \rangle+\sqrt{\langle n
\rangle}}+d^{2})^{\frac{1}{4}},r_{2c}=(c^{2}_{\langle n
\rangle-\sqrt{\langle n \rangle}}+d^{2})^{\frac{1}{4}},
\end{equation}
and
\begin{equation}
r_{1r}=(c^{2}_{\langle n \rangle}+d^{2})^{\frac{1}{4
}},r_{2r}=(c^{2}_{\langle n \rangle-1}+d^{2})^{\frac{1}{4}},
\end{equation}
with
\begin{equation}
\cos(\varphi_{1c})=(1+\frac{d^{2}}{c^{2}_{\langle n
\rangle+\sqrt{\langle n \rangle}
}})^{-\frac{1}{2}},\cos(\varphi_{2c})=(1+\frac{d^{2}}{c^{2}_{\langle
n \rangle-\sqrt{\langle n \rangle} }})^{-\frac{1}{2}},
\end{equation}
and
\begin{equation}
\cos(\varphi_{1r})=(1+\frac{d^{2}}{c^{2}_{\langle n \rangle}
})^{-\frac{1}{2}},\cos(\varphi_{2r})=(1+\frac{d^{2}}{c^{2}_{\langle
n \rangle-1}})^{-\frac{1}{2}}.
\end{equation}
From (), () and () we can see that gravitational field affect the
collapse and revival times. Moreover, we obtain the collapse and
revival times by $\lambda t_{c}=4.3$ and $\lambda t_{r}=1.9$,
respectively, with the same corresponding data used in the
sub-section 4a. Therefore, the collapse and revival times that we
have obtained from () and ()
are the same with the collapse and revival times that we have shown in figure 2a.\\
\\
{\bf  4c. Atomic momentum diffusion} \\
 \\
 The next quantity is the atomic momentum diffusion which is given
 by
\begin{equation}
\Delta p=(\langle \hat{p}^{2}\rangle - \langle \hat{p} \rangle^{2}
)^{\frac{1}{2}}.
\end{equation}
By using (18) and $\hat{p}|p\rangle=p|p\rangle$, we obtain
\begin{equation}
\Delta p=\{[\sum_{n=0}^{\infty}\int  d^{3}p p^{2}
(|\psi_{1}|^{2}+|\psi_{2}|^{2})  ] -[\sum_{n=0}^{\infty}\int
d^{3}p p (|\psi_{1}|^{2}+|\psi_{2}|^{2}) ]^{2} \}^{\frac{1}{2}}.
\end{equation}
Now we substitute (27) and (28) into (53) with
 assumes which is used in the sub-section 4a so that we can
obtain
\begin{eqnarray}
\Delta p=&&\{  [  \sum_{n=0}^{\infty}p_{0}^{2}(
|C_{1,\vec{p}_{0}}|^{2}[
|H(A_{n}(\vec{p}_{0}),B_{t}(\vec{p}_{0}))|^{2} \\
\nonumber + && |H(A_{n}(\vec{p}_{0})+1,B_{t}(\vec{p}_{0}))|^{2} ]+
|C_{2,\vec{p}_{0}}|^{2}[|_{1}F_{1}(-A_{n}(\vec{p}_{0}),\frac{1}{2};B^{2}_{t}(\vec{p}_{0}))
|^{2}\\
\nonumber   + &&| _{1}F_{1}(- \frac{1}{2}(
A_{n}(\vec{p}_{0})+1),\frac{1}{2};B^{2}_{t}(\vec{p}_{0}))|^{2}] +
2Re[C_{1,\vec{p}_{0}}C^{*}_{2,\vec{p}_{0}}\\
\nonumber &&(H(A_{n}(\vec{p}_{0})+1,B_{t}(\vec{p}_{0}))
_{1}F^{*}_{1}(-A_{n}(\vec{p}_{0}),\frac{1}{2};B^{2}_{t}(\vec{p}_{0}))\\
\nonumber  +&& H(A_{n}(\vec{p}_{0})+1,B_{t}(\vec{p}_{0}))
_{1}F^{*}_{1}(- \frac{1}{2}(
A_{n}(\vec{p}_{0})+1),\frac{1}{2};B^{2}_{t}(\vec{p}_{0}))  ) ] )] \\
\nonumber -&&[\sum_{n=0}^{\infty}p_{0}( |C_{1,\vec{p}_{0}}|^{2}[
|H(A_{n}(\vec{p}_{0}),B_{t}(\vec{p}_{0}))|^{2} \\
\nonumber + && |H(A_{n}(\vec{p}_{0})+1,B_{t}(\vec{p}_{0}))|^{2} ]+
|C_{2,\vec{p}_{0}}|^{2}[|_{1}F_{1}(-A_{n}(\vec{p}_{0}),\frac{1}{2};B^{2}_{t}(\vec{p}_{0}))
|^{2}\\
\nonumber   + &&| _{1}F_{1}(- \frac{1}{2}(
A_{n}(\vec{p}_{0})+1),\frac{1}{2};B^{2}_{t}(\vec{p}_{0}))|^{2}] +
2Re[C_{1,\vec{p}_{0}}C^{*}_{2,\vec{p}_{0}}\\
\nonumber &&(H(A_{n}(\vec{p}_{0})+1,B_{t}(\vec{p}_{0}))
_{1}F^{*}_{1}(-A_{n}(\vec{p}_{0}),\frac{1}{2};B^{2}_{t}(\vec{p}_{0}))\\
\nonumber  +&& H(A_{n}(\vec{p}_{0})+1,B_{t}(\vec{p}_{0}))
_{1}F^{*}_{1}(- \frac{1}{2}(
A_{n}(\vec{p}_{0})+1),\frac{1}{2};B^{2}_{t}(\vec{p}_{0}))  )])
]^{2}\}^{\frac{1}{2}},
\end{eqnarray}
where we apply all assumes in the sub-section 4a.\\
  \\
{\bf  4d. The Probability distribution of the cavity-field} \\
 \\
The probability distribution function $p(n)$ that there are n
photons in the cavity-field is given by
\begin{equation}
P(n)=|\langle n|\psi(t)\rangle|^{2}.
\end{equation}
By using the expressions (27) and (28) which represent the
probability amplitudes, we have
\begin{equation}
P(n)=\int d^{3}p [|\psi_{1}|^{2}+|\psi_{2}|^{2}].
\end{equation}
 Therefore, by assumes which is used in the sub-section 4a, we obtain the probability distribution
function $p(n)$ at time $t=\tau$
\begin{eqnarray}
p(n)=&& \{|C_{1,\vec{p}_{0}}|^{2}[
|H(A_{n}(\vec{p}_{0}),B_{\tau}(\vec{p}_{0}))|^{2} \\
\nonumber + &&
|H(A_{n}(\vec{p}_{0})+1,B_{\tau}(\vec{p}_{0}))|^{2} ]+
|C_{2,\vec{p}_{0}}|^{2}[|_{1}F_{1}(-A_{n}(\vec{p}_{0}),\frac{1}{2};B^{2}_{\tau}(\vec{p}_{0}))
|^{2}\\
\nonumber   + &&| _{1}F_{1}(- \frac{1}{2}(
A_{n}(\vec{p}_{0})+1),\frac{1}{2};B^{2}_{\tau}(\vec{p}_{0}))|^{2}]
+
2Re[C_{1,\vec{p}_{0}}C^{*}_{2,\vec{p}_{0}}\\
\nonumber &&(H(A_{n}(\vec{p}_{0})+1,B_{\tau}(\vec{p}_{0}))
_{1}F^{*}_{1}(-A_{n}(\vec{p}_{0}),\frac{1}{2};B^{2}_{\tau}(\vec{p}_{0}))\\
\nonumber  +&& H(A_{n}(\vec{p}_{0})+1,B_{\tau}(\vec{p}_{0}))
_{1}F^{*}_{1}(- \frac{1}{2}(
A_{n}(\vec{p}_{0})+1),\frac{1}{2};B^{2}_{\tau}(\vec{p}_{0}))  ) ]
\},
\end{eqnarray}
where we have introduced the functions in terms of $\vec{p}_{0}$
in the sub-section 4a.
  Moreover, in figures 3a and 3b we
consider the same corresponding data, respectively, used in
figures 1a and 1b with $\alpha=2$,$\tau=1.4\times10^{-6}sec$.
With comparing figures 3a and 3b we may see that the
gravitational field affect in
the probability distribution of the cavity-field.\\
 \\
{\bf  4e. Photon Counting Statistics} \\
\\
 We investigate another parameter for field
so-called Mandel parameter $Q(t)$ [39,40]
\begin{equation}
Q(t)=\frac{(\langle n(t)^{2}\rangle-\langle n(t)\rangle
^{2})}{\langle n(t)\rangle}-1,
\end{equation}
where $Q=0$, $Q<0$ and $Q>0$  for field show Poissonian,
sub-Poissonian and super-Poissonian statistic, respectively. We
define $\langle n(t)\rangle=\sum_{n=0}^{\infty}n(t)P(n)$ and we
have
\begin{equation}
Q(t)=(\{[\sum_{n=0}^{\infty}n^{2}(t)P(n)
]-[\sum_{n=0}^{\infty}n(t)P(n) ]^{2}    \}
[\sum_{n=0}^{\infty}n(t)P(n) ]^{-1})-1.
\end{equation}
Therefore, by the assumes which are used in sub-sections 4a and 4d we obtain\\
\\
\\
\begin{eqnarray}
Q=&&(\{[\sum_{n=0}^{\infty}n^{2}\{|C_{1,\vec{p}_{0}}|^{2}[
|H(A_{n}(\vec{p}_{0}),B_{t}(\vec{p}_{0}))|^{2} \\
\nonumber + && |H(A_{n}(\vec{p}_{0})+1,B_{t}(\vec{p}_{0}))|^{2} ]+
|C_{2,\vec{p}_{0}}|^{2}[|_{1}F_{1}(-A_{n}(\vec{p}_{0}),\frac{1}{2};B^{2}_{t}(\vec{p}_{0}))
|^{2}\\
\nonumber   + &&| _{1}F_{1}(- \frac{1}{2}(
A_{n}(\vec{p}_{0})+1),\frac{1}{2};B^{2}_{t}(\vec{p}_{0}))|^{2}] +
2Re[C_{1,\vec{p}_{0}}C^{*}_{2,\vec{p}_{0}}\\
\nonumber &&(H(A_{n}(\vec{p}_{0})+1,B_{t}(\vec{p}_{0}))
_{1}F^{*}_{1}(-A_{n}(\vec{p}_{0}),\frac{1}{2};B^{2}_{t}(\vec{p}_{0}))\\
\nonumber  +&& H(A_{n}(\vec{p}_{0})+1,B_{t}(\vec{p}_{0}))
_{1}F^{*}_{1}(- \frac{1}{2}(
A_{n}(\vec{p}_{0})+1),\frac{1}{2};B^{2}_{t}(\vec{p}_{0}))  ) ]
\}  ] \\
\nonumber -&&  [\sum_{n=0}^{\infty} \{|C_{1,\vec{p}_{0}}|^{2}[
|H(A_{n}(\vec{p}_{0}),B_{t}(\vec{p}_{0}))|^{2} \\
\nonumber + && |H(A_{n}(\vec{p}_{0})+1,B_{t}(\vec{p}_{0}))|^{2} ]+
|C_{2,\vec{p}_{0}}|^{2}[|_{1}F_{1}(-A_{n}(\vec{p}_{0}),\frac{1}{2};B^{2}_{t}(\vec{p}_{0}))
|^{2}\\
\nonumber   + &&| _{1}F_{1}(- \frac{1}{2}(
A_{n}(\vec{p}_{0})+1),\frac{1}{2};B^{2}_{t}(\vec{p}_{0}))|^{2}] +
2Re[C_{1,\vec{p}_{0}}C^{*}_{2,\vec{p}_{0}}\\
\nonumber &&(H(A_{n}(\vec{p}_{0})+1,B_{t}(\vec{p}_{0}))
_{1}F^{*}_{1}(-A_{n}(\vec{p}_{0}),\frac{1}{2};B^{2}_{t}(\vec{p}_{0}))\\
\nonumber  +&& H(A_{n}(\vec{p}_{0})+1,B_{t}(\vec{p}_{0}))
_{1}F^{*}_{1}(- \frac{1}{2}(
A_{n}(\vec{p}_{0})+1),\frac{1}{2};B^{2}_{t}(\vec{p}_{0}))  ) ]
\}  ]^{2}    \} \\
\nonumber &&[\sum_{n=0}^{\infty}\{|C_{1,\vec{p}_{0}}|^{2}[
|H(A_{n}(\vec{p}_{0}),B_{t}(\vec{p}_{0}))|^{2} \\
\nonumber + && |H(A_{n}(\vec{p}_{0})+1,B_{t}(\vec{p}_{0}))|^{2} ]+
|C_{2,\vec{p}_{0}}|^{2}[|_{1}F_{1}(-A_{n}(\vec{p}_{0}),\frac{1}{2};B^{2}_{t}(\vec{p}_{0}))
|^{2}\\
\nonumber   + &&| _{1}F_{1}(- \frac{1}{2}(
A_{n}(\vec{p}_{0})+1),\frac{1}{2};B^{2}_{t}(\vec{p}_{0}))|^{2}] +
2Re[C_{1,\vec{p}_{0}}C^{*}_{2,\vec{p}_{0}}\\
\nonumber &&(H(A_{n}(\vec{p}_{0})+1,B_{t}(\vec{p}_{0}))
_{1}F^{*}_{1}(-A_{n}(\vec{p}_{0}),\frac{1}{2};B^{2}_{t}(\vec{p}_{0}))\\
\nonumber  +&& H(A_{n}(\vec{p}_{0})+1,B_{t}(\vec{p}_{0}))
_{1}F^{*}_{1}(- \frac{1}{2}(
A_{n}(\vec{p}_{0})+1),\frac{1}{2};B^{2}_{t}(\vec{p}_{0}))  ) ]
\}   ]^{-1})-1,
\end{eqnarray}
where we have defined the functions in terms of $\vec{p}_{0}$ in
the sub-section 4a.
 In figure 4a with an initial coherent state for field and the same
data used in figure 1a we can see the mandel parameter until
second order in presence of the gravitational field is negative.
Therefore, the statistics is sub-Poissonian. Moreovere,  in
$\lambda t>0.7$,   the mandel parameter increases and in $\lambda
t=0.77$, this parameter is minimum. In figure 4b the mandel
parameter in  $\lambda t>0.7$ decreases when
$\vec{q}.\vec{g}$ is very small.\\
\\
\\
\\
{\bf  4f. Quadrature Squeezing of the Cavity-Field} \\
\\
 Now we investigate the quadrature squeezing of the radiation field [41-45] in the presence of
gravitational field. We introduce the Hermitian amplitude
operators
\begin{equation}
\hat{X}_{1}=\frac{1}{2}(\hat{a}+\hat{a}^{\dagger}),\hat{X}_{2}=\frac{1}{2i}(\hat{a}-\hat{a}^{\dagger}),
\end{equation}
where $\hat{a}$ and $\hat{a}^{\dagger}$ obey the communication
relation $[ \hat{a},\hat{a}^{\dagger}  ]=1$. A squeezed state of
the radiation field is obtained if
\begin{equation}
\langle(\triangle \hat{X}_{i})^{2}\rangle<\frac{1}{4},(i=1 or 2),
\end{equation}
where
\begin{equation}
\langle(\triangle
\hat{X}_{i})^{2}\rangle=\langle\hat{X}_{i}^{2}\rangle-\langle\hat{X}_{i}\rangle^{2}.
\end{equation}
The degree of squeezing can be measured by the squeezing
parameter $S_{i}, (i=1 or 2)$ defined by
\begin{equation}
S_{i}=4\langle(\triangle \hat{X}_{i})^{2}\rangle-1.
\end{equation}
 Therefore, from (18),(27) and (28) we obtain the
 squeezing parameter $S_{i}, (i=1 or 2)$
\begin{equation}
S_{1}=(\langle \hat{a}^{2} \rangle-\langle
\hat{a}\rangle^{2})+(\langle \hat{a}^{\dagger2} \rangle-\langle
\hat{a^{\dagger}}\rangle^{2})+2( \langle \hat{ a} ^{\dagger}
\hat{a} \rangle-\langle \hat{a} ^{\dagger}\rangle \langle\hat{a}
\rangle),
\end{equation}
and
\begin{equation}
S_{2}=-(\langle \hat{a}^{2} \rangle-\langle
\hat{a}\rangle^{2})-(\langle \hat{a}^{\dagger2} \rangle-\langle
\hat{a^{\dagger}}\rangle^{2})+2( \langle \hat{ a} ^{\dagger}
\hat{a} \rangle-\langle \hat{a} ^{\dagger}\rangle \langle\hat{a}
\rangle),
\end{equation}
 where
\begin{equation}
\langle \hat{a}\rangle= \int d^{3}p
\sum_{n=0}^{\infty}(\sqrt{n}\psi_{1n}\psi_{1(n-1)}^{*}+\sqrt{n+1}\psi_{2n}\psi_{2(n-1)}^{*}),
\end{equation}
\begin{equation}
\langle \hat{a ^{\dagger} }\rangle=\int d^{3}p
\sum_{n=0}^{\infty}(\sqrt{n+1}\psi_{1n}\psi_{1(n+1)}^{*}+\sqrt{n+2}\psi_{2n}\psi_{2(n+1)}^{*}),
\end{equation}
and
\begin{equation}
\langle \hat{a} ^{2} \rangle= \int d^{3}p
\sum_{n=0}^{\infty}(\sqrt{n(n-1)}\psi_{1n}\psi_{1(n-2)}^{*}+\sqrt{n(n+1)}\psi_{2n}\psi_{2(n-2)}^{*}),
\end{equation}
\begin{equation}
\langle \hat{a} ^{ \dagger2} \rangle= \int d^{3}p
\sum_{n=0}^{\infty}(\sqrt{(n+1)(n+2)}\psi_{1n}\psi_{1(n+2)}^{*}+\sqrt{(n+2)(n+3)}\psi_{2n}\psi_{2(n+3)}^{*}),
\end{equation}
with
\begin{equation}
\langle \hat{ a} ^{\dagger} \hat{a} \rangle= \int d^{3}p
\sum_{n=0}^{\infty}(n\psi_{1n}\psi_{1n}^{*}+(n+1)\psi_{2n}\psi_{2(n-1)}^{*}),
\end{equation}
where we define from (27) and (28) $\psi_{1n}=\psi_{1}(t)$ and
$\psi_{2n}=\psi_{2}(t)$, respectively, and from sub-section 4a we
apply all assumes and the initial conditions.
 In figures 5a and 5b we have plotted
the squeezing parameters $S_{i}, (i=1 or 2)$ versus the scaled
time $\lambda t$ for the same corresponding data, respectively,
used in figures 1a and 1b. As it is seen, each of the two
quadrature components exhibits squeezing in the course of time
evolution. Because of the influence of gravitational field, each
of the two quadrature components show oscillatory behavior.
\section{Summary and conclusions}
 We have studied the temporal evolution of quantum statistical properties of an
interacting atom-field system in the presence of a homogeneous
gravitational field within the framework of the Jaynes-Cummings
model. For this purpose, taking into account both the atomic
motion and gravitational field a full quantum treatment of the
internal and external dynamics of the atom has presented based on
an alternative su(2) dynamical algebraic structure. By solving
analytically the Schr\"{o}dinger equation in the interaction
picture, the evolving state of the system has found by which the
influence of the gravitational field on
 the dynamical behavior of the atom-field system has explored. Assuming that initially the
field has prepared in a coherent state and the two-level atom has
prepared in a coherent superposition of the excited and ground
states, the influence of gravity on the atomic dipole moment,
collapses and revivals of the atomic motion, atomic momentum
diffusion, photon counting statistics and quadrature squeezing of
the radiation
field has studied.  \\
\\
\\
{\bf  Acknowledgements} \\
On of the authors (M.M) wishes to thank The Office of Graduate
Studies of the Science and Research Campus Islamic Azad
University of Tehran for their support.

\vspace{20cm}
% ==================================================================================

{\bf FIGURE CAPTIONS:}

{\bf FIG. 1-4 } The dipole moment evolution versus the scaled
time $\lambda t$. Here
$q=10^{7}m^{-1}$,\\
$p_{0}=10^{-26}\frac{Kg.m}{s}$,$g=9.8\frac{m}{s^{2}}$,$\omega_{rec}=.5\times10^{6}\frac{rad}{s}$,\\$\lambda=9.7\times10^{6}\frac{rad}{s}$,$\triangle_{0}=8.5\times10^{7}\frac{rad}{s}$,
 $ \varphi=0$ and $c_{e}=c_{g}=\frac{1}{\sqrt{2}}$ with coherent state for initial cavity-field;\\

 {\bf a)}In the presence of gravitational field.

{\bf b)}For $\vec{q}.\vec{g}=0$.\\

{\bf FIG. 5 } The squeezing parameters versus the scaled time
$\lambda t$ with the same corresponding data
 used in fig.1-4;\\

{\bf a)} The squeezing parameter $S_{1}$ in the  presence of
gravitational field.\\

{\bf b)} The squeezing parameter $S_{2}$ in the  presence of
gravitational field.\\

%\begin{figure}[]

%\epsfxsize=12cm\centerline{\epsffile{colors.eps}}
%\includegraphics[angle=0,scale=0.7]{p2-1a.eps}
 % \caption[]{fig-a1}\label{a1}
%\end{figure}

\end{document}